\documentclass[twocolumn,amsmath,amssymb]{revtex4}
\usepackage{graphicx}% Include figure files
\usepackage{dcolumn}% Align table columns on decimal point
\usepackage{bm}% bold math

\begin{document}
\title{Comment on `` Signatures of Surface States in Bismuth at High Magnetic Fields ''}
\date{\today}
\maketitle

Nernst effect is the  transverse electric field due to a longitudinal thermal gradient in presence of magnetic field. Its magnitude was recently studied up to 33 T in a bismuth single crystal with a magnetic field oriented along the trigonal axis \cite{behnia1}. In addition to quantum oscillations of the Nernst response , a rich structure with several additional peaks was also detected when the magnetic field exceeded the quantum limit, which is of the order of 10 T in this configuration.  The origin of these high-field Nernst anomalies is a subject of ongoing research. Very recently,  Seradjeh, Wu, and Phillips\cite{seradjeh} have suggested that a surface state can give rise to these anomalies. This  proposal does not resist a critical examination.

Fig. 1 sketches the experiment\cite{behnia1}  designed to measure the Nernst response of a bulk single crystal. One can imagine cases in which the presence of a different state within a depth of $e$ would contaminate the outcome of such an experiment. For example, in the case of a resistivity experiment, if the state in the red region happens to be a superconductor, the  electric field in the bulk is expected to be short-circuited by the  presence of the superconductor, no matter how small $e$ is. In the case of a Nernst experiment on bulk bismuth, on the other hand, crude numbers are such that a contamination is unlikely, even for an implausibly large $e$.

Seradjeh \emph{et al.} \cite{seradjeh} write: ``The  samples in the experiment have a thickness of 0.8 mm; \emph{so despite their higher relative carrier density} [emphasize added], one expects the Nernst signal of the surface to be weaker than that of the bulk by a factor of at least 1000. This is certainly true  for fields below the quantum limit. However, above the quantum limit, the strong peaks from the bulk will be absent and the surface signal could be detected much more easily.'' The authors assume a deepness for the surface state as long as  $e= \sim$ 1 $\mu m$ (equivalent to 800 interatomic distances and 20 electron Fermi wavelengths) and take the width of the sample for its thickness. But the most fundamental mistake is elsewhere. The order of magnitude of the Nernst response in a given metal \emph{enhances} as the carrier density decreases and also as the carrier mobility increases. This conclusion is based on available Nernst data in various metals across seven orders of magnitude and directly pops out of the semiclassical picture of electron transport\cite{behnia2}. The magnitude of the Nernst response in bulk bismuth is unrivalled among metals, precisely because of its  exceptionally low  carrier density and its exceptionally high electronic mobility. The surface state, on the other hand, not only presents a much higher carrier density, but also has a lower electronic mobility (a feature  invoked by the authors to explain the absence of  `` surface quantum oscillations'' below 9 T). Therefore, \emph{the Nernst response of the surface state is expected to be a thousand times lower than bulk, even if, implausibly, it extends $e\sim$ 1 mm beneath the surface}. The non-oscillating background Nernst response of bulk bismuth, even above the quantum limit, is large enough ($\sim1 mVK^{-1}$ at 10 T and 1.5 K) to be many orders of magnitude beyond any contamination from the surface state.

In  this context, it is instructive to note that Nernst effect in bulk graphite\cite{zhu} is almost two orders of magnitude larger than in monolayer graphene\cite{zuev} and the profile of the quantum oscillations is qualitatively different in the two cases\cite{zhu}. In the two-dimensional case, a quantum oscillation of the Nernst response symmetrically sandwiches a vanishing signal , but in the three-dimensional case it presents an asymmetric peak. The Nernst peaks in bismuth, both below and above the quantum limit, present an asymmetric profile in a manner analogue to graphite\cite{zhu}.

 In summary, Seradjeh, Wu and Phillips leap across  many orders of magnitude to embrace their conclusions. This tends to disqualify their paper as a serious contribution to the ongoing debate on the origin of the high-field Nernst anomalies in bismuth.

\textbf{Kamran Behnia}\\
Laboratoire Photons Et Mati\'ere (CNRS-UPMC)\\  ESPCI,  Paris, France
\begin{figure}
  \resizebox{!}{0.22\textwidth}{\includegraphics{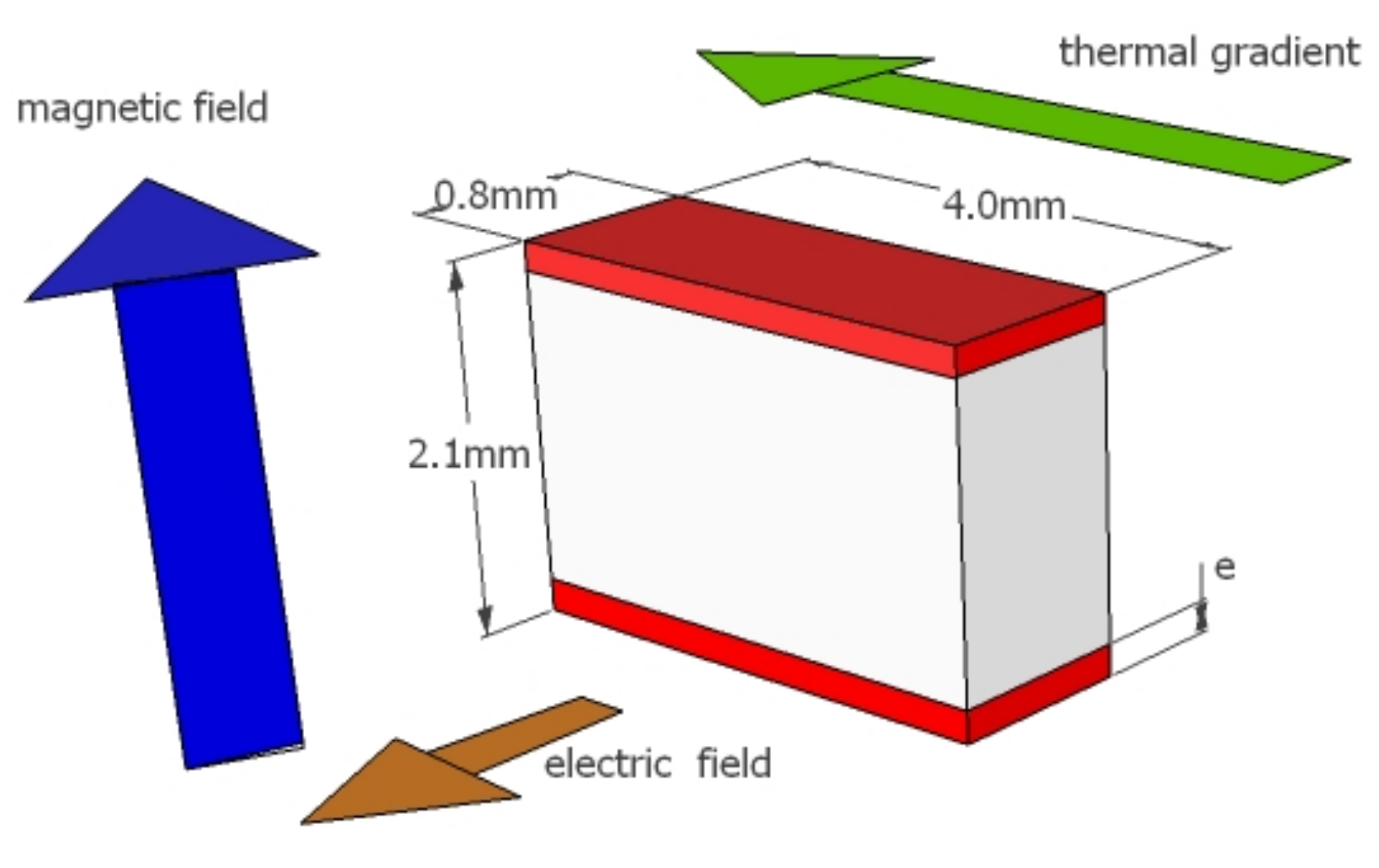}}
\caption{\label{Fig1} The configuration of the Nernst experiment with relevant dimensions. Seradjeh \emph{et al.}\cite{seradjeh} suggest that the measured voltage is dominated by the surface state in red.}
\end{figure}

\end{document}